# On the actual spatial resolution of Brillouin Imaging

S. Caponi [1], D. Fioretto [2], M. Mattarelli [2,*]

[1]*CNR-IOM - Istituto Officina dei Materiali - Research Unit in Perugia,*
*c/o Dep. of Physics and Geology, University of Perugia, Via A. Pascoli, I-06123 Perugia, Italy*
[2]*Dep. of Physics and Geology, University of Perugia, Via A. Pascoli, I-06123 Perugia, Italy*
*\*Corresponding author: maurizio.mattarelli@unipg.it*

**Brillouin imaging is an emerging optical elastography technique able to generate maps of the mechanical properties at microscale, with great potential in biophysical and biomedical fields. A key parameter is its spatial resolution, which is usually identified with that of the confocal microscope coupled to the Brillouin interferometer. Conversely, here we demonstrate that the mean free path of acoustic phonons plays a major role in defining the resolution, especially for high numerical aperture confocal setups. Surprisingly, the resolution of elastography maps may even deteriorate when decreasing the scattering volume.**

•https://doi.org/10.1364/OL.385072

In a peculiar manifestation of Heisenberg Uncertainty Principle, when focusing by a lens a collimated light beam, the increased spatial definition is obtained at the expense of the definition in the light wavevector [1,2]. This is usually of no concern in many microscopy techniques where the signal is scarcely affected by the direction of the exciting light, and even near-field excitation, with undefined light wavevector, can be effectively exploited[3]. Unfortunately, this is not the case for Brillouin Spectroscopy (BS). As a matter of fact, in its standard configuration, BS is sensitive to spontaneous acoustic vibrations which are thermally activated in any material. Among all these vibrations, called "phonons", the scattering configuration (which includes wavevector and polarization of the exciting and collected light) selects a subset of phonons whose properties are reflected on the measured spectrum. In particular, in the usual description of a homogeneous visco-elastic medium, Brillouin spectra present a peak at frequency shift $\omega_B$ and width $2\Gamma_b$ for each exchanged wavevector **q**. These parameters are related to the real and imaginary parts of the longitudinal elastic modulus through the following relationships: $M' = \rho \omega_B^2 / q^2$ and $M'' = \omega_B \Gamma_b \rho / q^2$, being $\rho$ the density of the medium (details can be found in recent reviews, e.g. [4–6]). More complex is the case of heterogeneous media. In this case, changes in the elastic modulus within the scattering volume give rise to Brillouin peaks at different frequencies and/or to heterogeneous broadening of Brillouin peaks. Decreasing the dimension of the scattering volume seems to be the most appropriate solution to improve the resolution of Brillouin elastography maps. Following this idea, in the last years, the challenging characterization of mechanical proprieties at the microscale [7,8] stimulated the development of confocal Brillouin microscopes[9–14], where the use of high NA objectives allows the reduction of the scattering volume down to the sub-micrometer scale and allows effective depth-scans of transparent samples, as demonstrated in cornea [15][16]. Micro-Brillouin spectroscopy revealed mechanical modulations in subcellular compartments[10,17–19] and in biological tissues[13,20,21]. However, increasing the numerical aperture increases the range of detected scattering wavevectors, with noticeable and counterintuitive effects in Brillouin spectra. For the first time in this letter, we face this issue providing evidence of the effect of the **q** angular spread on the spatial resolution of the Brillouin imaging. The present study has a fundamental as well as a technological relevance giving insights for correct data interpretation and for future instrumentation design.

Fig. 1 presents the calculated magnitude and angular distribution of the wavevectors collected by a micro-Brillouin experiment in two different "backscattering" configurations corresponding to the use of air objectives with low (0.3) and high (0.9) NA. In the computation, a homogeneous incoming beam covering all of the objective aperture and an isotropic scattered radiation were considered. Fig. 1 b), shows that the NA increase leads to a significant spread in |q|: using the 0.9 NA objective in backscattering configuration, the average value of |q| is about 10% lower than the nominal one. Moreover, as reported in Fig.1 c), NA hugely affects the angular distribution of the collected wavevectors: the deviation θ from the optical axis reaches an average value of 8.5° for the low NA objective and 32° for the high NA one .The **q**-distribution has two remarkable effects on the Brillouin spectra: on one hand, it lowers the spectral definition, due to the fact that both the frequency and the width of the peak depend on |q| [22–24] . In this case, a straightforward, though delicate, deconvolution procedure is needed to extract the relevant material parameters [17]. On the other hand, the **q**-distribution implies also a distribution in the investigated directions, which in NA=0.9 optics can reach up to 65° from the optical axis. This is not relevant in homogeneous isotropic samples, but it becomes a fundamental issue when analyzing heterogeneous samples, i.e. just the ones where Brillouin imaging is of interest.

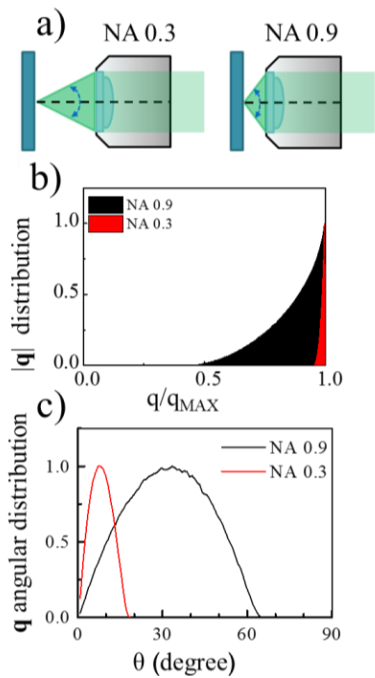

Fig. 1. q distribution of the collected signal for a Brillouin scattering process in two different "backscattering" configurations exemplified in the upper cartoons for two NAs. The red and black curves represent the q distribution of the modulus, b), and of the angular deviation from the optical axis, c), for NA=0.3 and NA=0.9, respectively.

In fact, given the propagating nature of the vibrations detected by BS, the **q** angular distribution translates into a lateral broadening of the investigated region. The extent of this effect can be a priori estimated considering the mean free path of acoustic phonons, $L_2$. (sketched in yellow in Fig. 2). In principle, in a lossless homogeneous material such length is infinite. In real systems $L_2$ gets finite because of dissipative mechanisms, be they due to static reasons, such as elastic heterogeneities on the scale of the phonon wavelength ($L_1=2\pi/q\sim200$ nm for 532 nm lasers in backscattering) or dynamic ones such as relaxation processes[25–28]. Note that both kind of mechanisms are present in biological materials and affect their Brillouin spectra [5,19,20,29]. However, BS itself provides a quite effective way to measure $L_2$. In fact, the peak width is inversely proportional to the phonon life time, and knowing the Brillouin peak position it is possible to obtain $L_2$ from the quality factor of the peak: $L_2 = \omega_B /2\Gamma\ L_1$ [30]. Just as an example, in pure water the propagation length $L_2$ of the phonons detected in backscattering configuration using a 532 nm laser beam is ∼ 2.1 μm.

A rough evaluation of the consequences of phonon propagation on micro-Brillouin spatial resolution can be obtained considering the broadening of the lateral resolution as arising from the average projection of $L_2$ by the off-axis angle, $<Sin(\theta)\ L_2>$. The actual Brillouin resolution will be then given by a spatial convolution of this contribution with the scattering volume individuated by the optical system. Always referring to water, the lateral broadening due to acoustic phonon propagation is 0.3 and 1.1 μm for the two angular distributions considered in Fig. 1 c). It is worth to notice that for the two considered NA, the Abbe diffraction limit is 0.7 and 0.2 μm, respectively, so that the acoustic contribution to spatial resolution for high NA optics would be the dominant one.

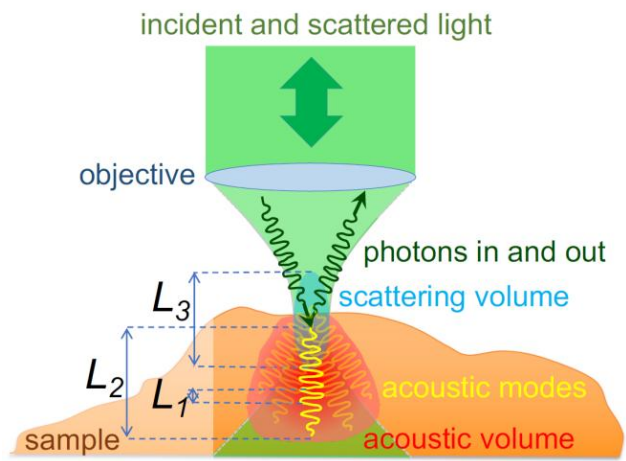

Fig. 2. Schematic micro-Brillouin scattering diagram. L1, L2 and L3 denote the relevant length scales for this interaction, given by phonon wavelength, mean free path and scattering volume, respectively. The acoustic volume, in red, takes into account the broadening induced by the phonon propagation.

In order to substantiate these considerations with experimental data, we measured the evolution of Brillouin spectra across a sharp interface between different materials. The change in intensity of the Brillouin peaks allows reconstructing the Edge Spread Function (ESF), whose derivative called Point Spread Function (PSF) provides the spatial resolution of the technique [17,21]. In particular, we analyzed two different interfaces: (i) between Polyethylene (PET) and Glycerol and (ii) between a sodium-silicate glass and water (Fig.3a). The choice of these two couples of materials was dictated by their relatively small refractive index mismatch and, more important, by their very different acoustic mismatch, so providing a direct verification of the effect of the phonon mean free path on the resolution. In fact, we can expect that for low acoustic mismatch, i.e in the PET-Glycerol system, the vibrational modes in the two materials can couple across the interface, whereas this effect is strongly hindered in the water-glass system. As the determination of the spatial resolution of the measurement can be affected by the sharpness of the interface, we carefully prepared the sample with optically polished interfaces, parallel to the optical axis. Moreover, in order to discriminate between optical and acoustic contributions to the ESF, we exploited the unique characteristics of our Raman-Brillouin combined set-up [14] and simultaneously measured the ESF by both Brillouin and Raman Spectra (RS). Raman spectroscopy in fact, being related to the detection of non-propagating optical phonons, can represent a good proxy for the optical contribution to the ESF as detected by the transition from one to the other material within the scattering volume. Two different objectives were used in order to change the excitation/collection configuration: a 20X with NA 0.42 and a 60X water immersion objective with 1.2 NA. Note that the angular spreads of these objectives, when evaluated in the materials under consideration, match those of the objectives NA=0.3 and NA=0.9 in air, previously analyzed in Fig. 1. Several linear scans through the interface were probed for the two samples. As an example, in Fig. 3 b) we show the evolution of Raman spectra acquired in the PET-Glycerol sample (Fig. 3 a) and Fig. 3 c) reports the Brillouin spectrum acquired at the interface together with its fitting curve.

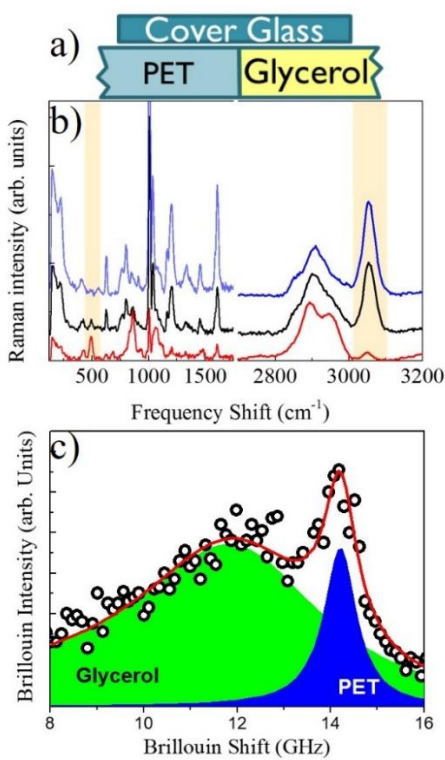

Fig. 3- a) Experimental configuration; b) Evolution of the Raman spectra through the PET-Glycerol interface. The peaks located at ~500 cm-1 and ~3050 cm-1 are respectively selected to evaluate the glycerol and the PET presence in the scattering volume; c) Typical Brillouin spectrum at the PET-Glycerol interface. The experimental points were fitted by two DHO functions, fixing the frequency and width of the PET Brillouin peak at 14.0 GHz and 0.8 GHz for PET, and at 12.0 GHz and 4.9 GHz for Glycerol. The full dataset consist of Raman and Brillouin spectra acquired with 0.3 or 0.5 μm steps for the glass water and the PET- glycerol interfaces, respectively. For the other interface, the frequency and width of the Brillouin peak is at 7.4 GHz and 0.9 GHz for H$_2$O, and at 32.3 GHz and 0.8 GHz for Glass.

In detail, the Brillouin spectra were analyzed using the sum of two DHO peaks as fit function, keeping fixed the position and width of the peaks as determined from measurements on pure materials, and allowing only the intensity to vary.

For the Raman spectra, few well-separated characteristic lines were used to estimate the materials presence. The intensity of the peaks, both Brillouin and Raman, changes continuously across the interface. Fig. 4 a) and Fig. 4 b) report the ESF obtained for the two investigated interfaces with the 60X objective. The ESF were fitted by an error function, having 2σ as width of the corresponding Gaussian PSF (see the inset of the figures).

Fig. 4 c) d) report the measured widths for the two couples of materials and for the two optical setups obtained from Brillouin, 2σ$_B$, and Raman, 2σ$_R$, measurements- The width of the laser spot, 2σ$_L$, as measured when focusing on an optically polished silicon wafer, is also indicated[17]. The Raman width 2σ$_R$ appears equal or slightly larger than the spot width, taking also into account the small effect at these length scales due to the rugosity of the sample surfaces. What is striking is the different behavior when changing the numerical aperture of the objective: while 2σ$_R$ follows strictly the laser spot size, the behavior of the Brillouin width 2σ$_B$ strongly depends on the couple of interfaced materials, increasing for the PET and Glycerol interface and decreasing for the water-glass one. This astonishing effect is the consequence of the extended nature of the phonons. In fact, at the interface the local phonon density of the states is modified, with hybridization between the modes of the two materials, broadening the effective interface to the region where the acoustic modes live [31][32] This effect, depending on the overlapping between the vibrational density of states, is negligible in the glass-water system as the difference in the Brillouin peak frequency ~25GHz is much larger than peak linewidths, but significant in the PET-Glycerol system, where the Brillouin peaks have a large region of superposition (Fig.2c). Therefore, even when the scattering volume is completely inside one component, the collective vibrational modes living across the interface can still be detected. A counterintuitive outcome of this analysis is the observation that, when reducing the scattering volume (passing from low to high NA), the lateral resolution in the measure of the elastic properties becomes worse (1.4 μm against 1.0 μm).

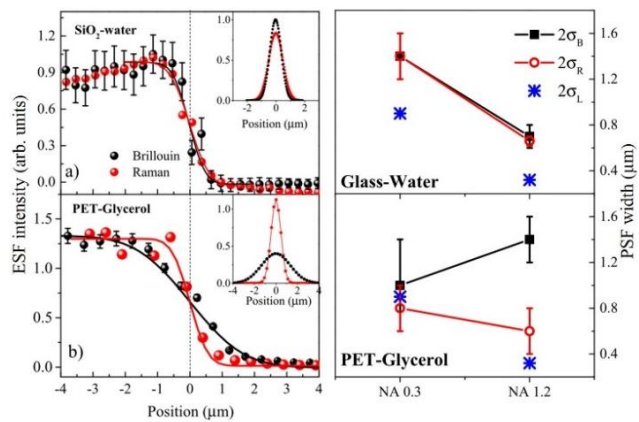

Fig. 4 Rescaled ESF measured for Raman (red) e Brillouin (black) experiments across the glass-water interface a) and across the PET-Glycerol interface b). The dots are the experimental data and the lines are the fits. In the insets: Derivative of the fitting functions providing

the PSF of the experimental setup. c) Width of the PSF (Brillouin, Raman, Si Reflectance indicated by black, red and blue dots, respectively) obtained from the two interfaces (Glass–Water and PET/Glycerol) in the two optical configurations (20X NA 0.3, 60X NA 1.2))

A further confirm can be obtained by the quantitative comparison between the experimental results and the previously reported calculations. In fact, the Brillouin resolution depends on the convolution of the scattering volume with the acoustic contribution, $2\sigma_E$, due to the acoustic phonon propagation. Using $2\sigma_R$ as lateral dimension of the scattering volume and considering as a first approximation that the contributions to the spatial resolution add in quadrature (exact for Gaussian distributions), the lateral acoustic contribution can be found as $2\sigma_E = 2\sqrt{(\sigma_B)^2 - (\sigma_R)^2}$. While this value is always utterly negligible for the glass-water interface, for the PET-Glycerol one studied by 1.2 NA objective $2\sigma_E$ results (1.2 ± 0.4) µm, indeed the dominating term in the final resolution. This value should be compared with the average radial projection of the mean free path $L_2$ along the angle θ. $L_2$ estimated from the Brillouin spectra as $L_2 = \omega_B/2\Gamma\ L_1$ is ~2.5 µm for PET and ~1.6 µm for Glycerol, while the average sin (θ) from the angular q distribution reported in Fig. 1 c) is ~0.52. The calculated value for $2\sigma_E$ thus falls in the range 0.8 – 1.3 µm, in good agreement with the experimental value.

The present analysis is of significant relevance for a number of recent works[17–19,21,33,34], where submicrometric features are discriminated on the basis of Brillouin signature. To this respect, it is worth noting that the observed elastic modulations are indeed related to genuine submicrometric structural changes, as suggested by comparison with different optical techniques. The present work just suggests that the measured values of the elastic constants should be considered as spatial averages, whose extent depends on both NA of the optics and on the propagation length of phonons, in turn depending on the investigated material. As a consequence, the optimization of the resolution could even require the adoption of lower NA optics. For example, in a recent work[21], a reduction in the mechanical contrast was observed passing from NA= 0.85 to NA=1.28. Our present findings suggest an influence in this result to come from the acoustic contribution, which, for the two objectives used in ref [21] is expected to be ~0.3 and ~0.6 µm, respectively.

Finally, as hinted by the measurements in the water-glass interface, it is important to highlight the strong influence of interfaces on the acoustic contribution. Taking it to the limit, in the case of complete phonon confinement, as occurs in nanoparticles, nanotubes or in films, the acoustic vibrations are standing waves in the confined directions. In this condition, the spatial resolution in the measurements of the elastic properties can be even higher than the optical resolution, also reaching the nanoscale [35–39].

In conclusion, we have shown that the ultimate resolution in the mechanical properties attainable by Brillouin imaging is limited not by the optical system but by the propagation properties of the investigated phonons. As a matter of fact, in Brillouin spectroscopy phonons and not light are the probe of the mechanical properties of the material. The light (according to the scattering geometry) only allows selecting the range of phonons by which we interrogate the material. The selection is in frequency by the |q| spread, in direction by the angle distribution of q, and in space by the scattering volume. However, the spatial resolution depends not only on the dimension of the scattering volume, but also on the mean free path $L_2$ of the acoustic phonons, which can extend over a much larger space. As a consequence, the actual elastic resolution can indeed get worse when increasing the optical resolution (the NA). Note that this is an inescapable constraint arising from the very fundamental reason that the detected excitations are extended over a finite space.

**Disclosures**. The authors declare no conflicts of interest.